\documentclass[journal abbreviation]{copernicus}

\usepackage{lineno}
\linenumbers

\begin{document}

\title{A method to constrain the characteristic angular size of the brightest cosmic-ray sources observed above
$57 \times 10^{18}$ eV}

\author{ }







\received{}
\pubdiscuss{} 
\revised{}
\accepted{}
\published{}


\firstpage{1}

\maketitle

\begin{abstract}
We introduce a method to constrain the characteristic angular size of the brightest cosmic-ray sources 
observed above $57 \times 10^{18}$ eV.
By angular size of a source, we mean the effective angular extent over 
which cosmic-rays from that source arrive at earth.
The method is based on 
the small-scale ($< 10^\circ$) self-clustering of cosmic-ray arrival directions.
The method is applicable to sparse data sets in which 
strong localizations of CR* directions are not yet observed.
We show that useful constraints on the source size can be made
in the near future and that these constraints are not strongly dependent
on the assumed
spatial distribution and luminosity function of the cosmic-ray sources.
We suggest that an indication of the source size is quite telling. 
For example, an indication of the source size can be
used to infer limits on the particle charge and intervening magnetic fields (not independently),
both of which are not well constrained so far. This is possible because the source size
is similar in scale to the magnetic deflection.
\end{abstract}


\introduction

We describe a new analysis method to constrain the characteristic angular size of the 
brightest cosmic-ray (CR) sources observed above $57 \times 10^{18}$ eV.
To facilitate our discussions, we use the symbol
CR* hereafter to denote cosmic-rays with energy greater than 
$57 \times 10^{18}~\mbox{eV}$.

We mean the angular size $s$ of a CR source to be the effective angular extent over which CR from
the source arrive at earth. A more rigorous definition is developed later.
We denote the angular size characteristic of the brightest CR* sources as $\bar{s}$.

Our method relies on two starting hypotheses: (1) the ultra-high energy cosmic-ray sources are
located in galaxies other than our own or neighbors closer than 1 Mpc, and (2) the
ultra-high energy cosmic-rays are protons or atomic nuclei that lose energy due
to interactions with the cosmic microwave background, i.e., the
Greisen-Zatsepin-Kuzmin (GZK) effect \citep{Greisen, Zatsepin}.
It has been shown \citep{Younk} that these hypotheses imply that
the fraction of flux $\bar{Q}$
from the brightest CR* source is at or above a few percent.
This result holds for a large range of the space number density $\rho$ of CR* sources.
For example, $\bar{Q}$ is proportional to the characteristic distance between sources (i.e., $\rho^{-1/3}$)~\citep{Younk} so that whether there is 1 observable CR* source
or 1000, we should expect $\bar{Q}$ to only change by a factor of 10.

The number of CR* observed by all experiments to date is $n_{obs} \approx 70$,
and this number is increasing by approximately 23 CR* per year \citep{Abraham2007}.
If hypotheses 1 and 2 are correct, it is likely that we will observe
in the near future several source pairs from the brightest sources, where a source pair is defined as two CR* that
originated from a common source.


The existence of source pairs
implies that the  clustering properties of CR*  arrival directions contain
information about the value of $\bar{s}$ .
In particular, the existence of source pairs will cause an increase in the observed
number of pairs with separation angles $\leq \bar{s}$.
In this way, the value of $\bar{s}$ affects
the shape of the 2-pt autocorrelation spectrum.

Our method uses a metric $m$ to quantify the shape of the
2-pt autocorrelation spectrum at  small angular scales
in order to explore the region $\bar{s} <  10^\circ$.
Using a Monte Carlo simulation built on hypotheses 1 and 2, we make predictions for $m$
based on the value of $\bar{s}$, the number $n_{obs}$ of observed CR*,
and the distribution of sources.
We show how a measurement of $m$ together with
these predictions of $m$ 
can be used to constrain $\bar{s}$.

Useful constraints are possible before 
strong localizations of CR* directions are observed
(i.e., before the value of 
$\bar{s}$ is trivially apparent). For example, if a lack of small scale clustering is observed, 
our method allows for a constraint such as $\bar{s} \geq 10^\circ$.
Indeed, this general idea has also been suggested by \citet{Cuoco}.


This work contrasts to many works (e.g., \citet{Nemmen}) that suggest limits on the magnetic deflection
of CR, in that here we do not require assumptions as to what objects accelerate ultra-high energy cosmic rays.
In this way, this work is somewhat similar to the work of \citet{Erdmann}, but the method described here is more general; e.g., we do not assume
details of how the clustering of events changes with energy.

This paper is organized as follows. In Section~2, we
consider how CR* from an extragalactic source may be distributed on the sky, and
we demonstrate the plausibility of $\bar{s} < 10^\circ$.
In Section~3, we present a
clustering metric $m$ based on the 2-point autocorrelation function
that is particularly sensitive to $\bar{s}$, while being rather insensitive to the number density $\rho$ of sources.
In Section~4, we describe a Monte Carlo algorithm used to predict $m$. In
Section~5, we show the predictions for $m$ and discuss how these predictions, together with an
observed value of $m$, 
can constrain $\bar{s}$ in the near future. We discuss how these constrains can
increase our understanding of the ultra-high energy cosmic-rays.
In Section~6, we conclude with a summary statement.

\section{CR* source morphology} \label{sec:morphology}

Let us consider how CR* from an extragalactic source may be distributed on the
sky. Imagine a source that emits protons isotropically from a point-like region.
The direction of this source is at a mid-galactic latitude, $B = -30^\circ$, and
$90^\circ$ from the galactic center, $L = -90^\circ$. The source is nearby
(i.e., it is one of the brightest sources in the sky) such that the injection
spectrum is not strongly modified by GZK energy losses. Therefore, an observed
spectrum of $dn/dE \propto E^{-2.6}$ with a maximum energy of $E_{max} = 3
\times 10^{20}~\mbox{eV}$ is plausible. This spectrum is similar to what has
been suggested by \citet{Allard}. For the regular field of the galaxy, we assume
BSS\_S symmetry and use the model described by \citet{Harari1999}, which is a
modified version of the model described by \citet{Stanev}. We assume that CR*
are not in the lensing regime of the turbulent component of the galactic
magnetic field. This has been suggested by \citet{Harari2002}. For this case,
the dispersion of CR* arrival directions by the turbulent component is less than
the dispersion by the regular component, and can be neglected. We assume the
dispersion of CR* arrival directions by extragalactic magnetic fields can also
be neglected. We take the detector resolution as $1^\circ$.

\begin{figure}[t]
\vspace*{2mm}
\begin{center}
\includegraphics[width=8.3cm]{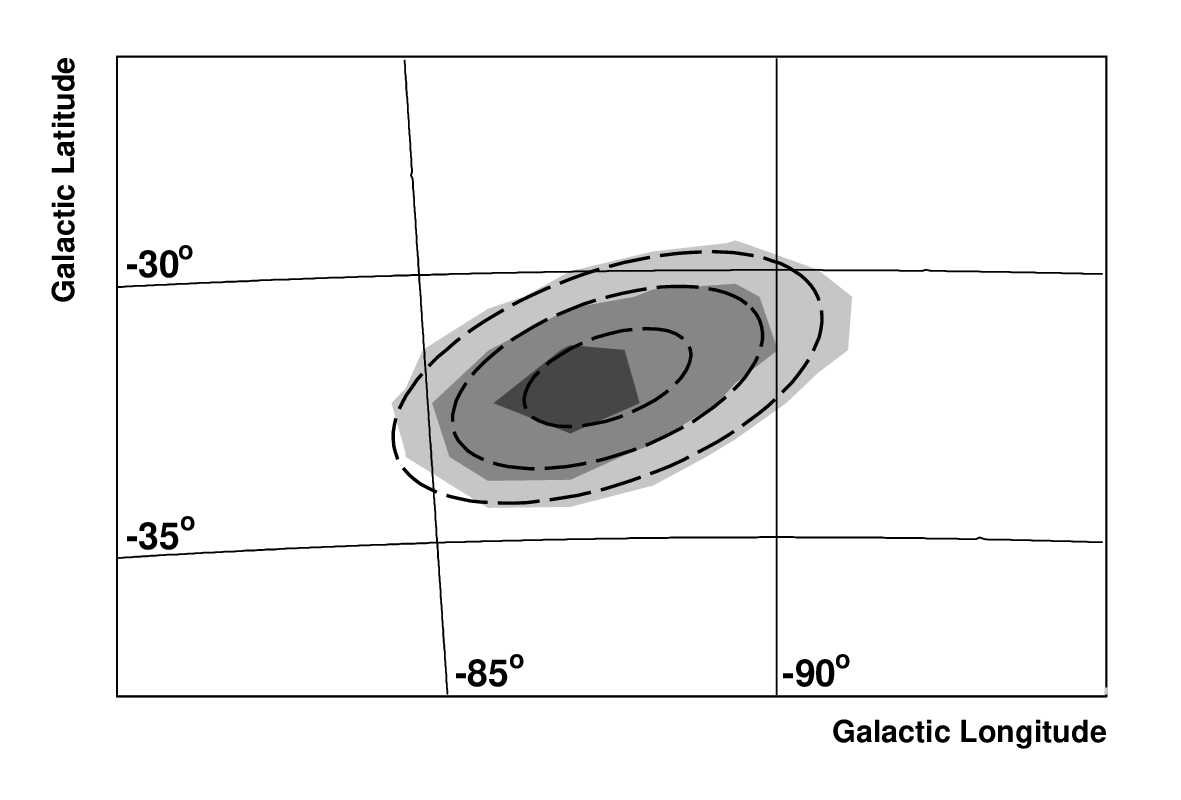}
\end{center}
\caption{Distribution of CR* from a simulated source. See text for details.}
\label{fig:SampleSource}
\end{figure}

In Fig.~\ref{fig:SampleSource}, we show a gnometic projection of the expected
CR* arrival directions from this source (i.e., the expected surface brightness).
The shading indicates three surface brightness contours: 70\%, 30\%, and 10\% of
the maximum surface brightness. 
Note that the location of the maximum surface brightness is offset from the
center of the distribution toward the low energy side
(i.e., further away from the actual source direction).  

Also in Fig.~\ref{fig:SampleSource}, we show an
elliptical Gaussian function fitted to the surface brightness distribution. The
dotted lines show the corresponding contours for this function. The center point
of the Gaussian function is located at $L = -87.9^\circ$ and $B = -32.0^\circ$. 
Thus, the characteristic magnetic deflection of CR* from this source is
approximately $2.9^\circ$, similar to the results of \citet{Harari1999}. 
The major axis is $49.5^\circ$ from north. In relation to the center point and
major and minor axes, the Gaussian function is described as
\begin{linenomath}
$$P(x,y) = A ~\mbox{exp} (-\frac{x^2}{2\sigma_x^2} -\frac{y^2}{2\sigma_y^2}),$$
\end{linenomath}
where $x$ is measured from the center point along the major axis, $y$ is
measured from the center point along the minor axis, $\sigma_x = 1.4^\circ$,
$\sigma_y = 0.8^\circ$, and $A$ is a normalization factor. Note that
the magnitude of $\sigma_y$ is similar to the angular resolution.

We define the source size to be 
\begin{linenomath}
$$s = 2 \sqrt{\sigma_x \sigma_y} = 2.1^\circ.$$
\end{linenomath}
For a source with $\sigma_x = \sigma_y$, approximately 86\% of the CR* are
observed within $s$ of the centroid. 
The source size $s$ can be thought of as a
first order structural term (i.e., it takes at least two CR* directions to
estimate it).

We define the source aspect ratio as 
\begin{linenomath}
$$\omega = \sigma_x / \sigma_y = 1.7.$$
\end{linenomath}
This can be thought of as a second order structural term (i.e., it takes at
least three CR* directions to estimate it).

For the actual CR* sources, the characteristic values for $s$ and $\omega$
depend on several details, many of which are not well constrained. For example,
simply changing the galactic latitude and longitude of a source can change $s$
by a factor of two. Here, we only wish to show that $\bar{s} <
10^\circ$ and $\omega \approx 1$ are plausible. That is, this parameter space is
worth investigating.


\section{Small-scale clustering metric}

We quantity the shape of the 2-pt autocorrelation function
with a clustering metric.
The amount of clustering $M$ at an angular scale
$\chi$ is quantified by the number of CR* pairs with angular separation less
than $\chi$, with each CR* pair weighted by the inverse of its angular
separation and by $1/\chi$. Symbolically,
\begin{linenomath}
$$ M(\chi) = \frac{1}{\chi} \sum^{n}_{i=2} \sum^{i-1}_{j=1} \Theta(\chi -
\beta_{ij})/\beta_{ij}, $$
\end{linenomath}
where $\beta_{ij}$ is the angular separation between CR* directions $i$ and $j$,
$\Theta$ is the step function, and $n_{obs}$ is the number of CR*. The
motivation to
weight each pair by the inverse of its angular separation comes from the fact
that for an isotropic distribution of CR*, the expected number of pairs with an
angular separation $\beta$ is $\left\langle dn_p/d\beta \right\rangle \propto
\beta$. This is valid for small $\beta$. We weight $M$ by $1/\chi$ so that
$\left\langle M(x) \right\rangle \approx \left\langle M(y) \right\rangle$ for an
isotropic distribution of CR*, where $x$ and $y$ are  small angles.

Then we define our clustering metric to be the ratio of the amount of
clustering at $2.5^\circ$ to the amount of clustering at $10^\circ$  
\begin{equation}
m = M(2.5^\circ)/M(10^\circ).
\label{eq:ClusteringMetric}
\end{equation}

With this definition, $0 < m < 4$. 
The lower limit results when all the pairs with separation angles less than
$10^\circ$ have  separation angles greater than $2.5^\circ$. In this case
$m = 0/M(10^\circ)$.
 The upper limit results when all the pairs with separation angles less than
$10^\circ$ have separation angles less than $2.5^\circ$.
In this case $m = 10^\circ / 2.5^\circ = 4$.
If $n_{obs}$ is too small, there is a possibility of $m =
0/0$. We work with data sets where $n_{obs}$ is large enough for this not to be a
concern.

The metric $m$ is a simple yet effective discriminator of different
$\bar{s}$ scenarios.
In Fig.~\ref{fig:AspectRatio} we show the value of $m$ as a function of
$s$ and $\omega$ for a single elliptical Gaussian source when $n_{obs}$ is large. 
The value of $m$ is strongly dependent on $s$ for $1^\circ < s <
10^\circ$, which is the parameter space we wish to explore. 
If we were interested in testing values of $\bar{s}$ greater than $10^\circ$, 
the pair of angles used in Eq. 1 would no longer be appropriate.
The value of $m$ is only slightly dependent on $\omega$ for $1 < \omega < 4$. For
interpreting results, a clustering metric that is only slightly dependent on
$\omega$ and other higher order structural terms is convenient.

Defining $m$ as a ratio of $M$ values makes our clustering metric indicative of the shape
of the autocorrelation spectrum at small angular scales
(i.e., where the feature created by $\bar{s}$ is located).
This is beneficial in constraining $\bar{s}$, especially in constraining $\bar{s}$ independent of $\rho$.
%
For example, the amount of clustering at $2.5^\circ$ relative to $10^\circ$ 
is strongly affected by $\bar{s}$ but not by $\rho$.
In contrast, the absolute amount of clustering at a single particular small angle (e.g., $M(2.5^\circ)$) is strongly affected by both
$\bar{s}$ and $\rho$.
In Section 5, we demonstrate the ability of the metric $m$ to discriminate between different $\bar{s}$ scenarios
independent of $\rho$.


By not including an energy term in $m$, our results do not depend on how the
morphology of the source or the apparent position of the source 
changes with threshold energy. Magnetic lensing effects (the
formation of multiple images \citep{Harari2002}) and the finite angular
resolution of the detector may make the CR* dispersion angle a complicated
function of energy. In particular, magnetic lensing effects are difficult to
predict because the magnetic field is not well known. 

%

%
%


\section{Monte Carlo Algorithm}

Our algorithm generates sets of CR* arrival directions given $n_{obs}$,
$\bar{s}$, and
$\rho$. Model details are based on hypotheses 1 and 2 from Section 1. An expected range of $m$
is calculated for different sets of input parameters.

\subsection{Source Distribution Models}

It is expected that the actual distribution of CR* sources is related in some
way to
the distribution of galaxies, but the details are not known. For example, we
have only broad constraints on the luminosity function of CR* sources, and we do
not
know in what environments the host galaxies are preferentially found (e.g.,
clusters or groups). To see how these details affect our results, we test
several different scenarios.

We test different luminosity functions by assuming each source is equally
luminous and then scanning over a large range of $\rho$. 
Including another free parameter (e.g., the shape or break-point of the
luminosity function) does not significantly improve the simulation.
In this case, $\rho$ does not represent 
the number density of all sources. Instead $\rho$ represents
the number density of a sub-set of sources that produces the majority of flux
(e.g., $\rho$ would not include a low luminosity tail).
We expect to more rigorously show the affects of  
scanning over different luminosity functions in further work.

We scan over the plausible range $10^{-6}~\mbox{Mpc}^{-3} \leq \rho \leq
10^{-3}~\mbox{Mpc}^{-3}$.

The lower limit for $\rho$ is chosen to be consistent with the observations of
cosmic-rays with energy $E > 10^{20}~\mbox{eV}$ and our postulate of GZK energy
losses. Above $10^{20}$ eV, the energy loss length for protons and iron-like
nuclei is only tens of Mpc \citep{Harari2006}. In this same energy range, the
energy loss lengths of intermediate weight nuclei are much less than either
protons or iron-like nuclei. Then if the CR* are baryonic, it is likely that
they are predominately protons or iron-like nuclei and that there are at least a
few CR* sources within 100 Mpc.

The upper limit for $\rho$ is chosen to be consistent with our postulate that no
CR* sources (including sources in the low luminosity tail) are located in the
Milky Way and its closest neighbors. The number density of galaxies with
luminosity $L>L^*$ (i.e., large galaxies) is approximately
$10^{-3}~\mbox{Mpc}^{-3}$ \citep{Liske}.

We consider two simple yet highly contrasting models for how the sources are
correlated with galaxies. In the first model, the sources are distributed evenly
(i.e., every location has equal probability of containing a source) except that
no source is allowed at a distance $d < 1$ Mpc. In the second model, the sources
are distributed proportional to the distribution of large galaxies out to 60 Mpc
and evenly distributed at greater distances. The cut at 60 Mpc facilitates the
construction of a volume-limited sample of large galaxies, and is justified in
that most of the structure in source directions must occur at small source
distances (i.e., the characteristic size of super clusters is a few tens of
Mpc.)

To construct a volume-limited sample of large galaxies, we use the PSCz catalog
\citep{Sanders}. The PSCz catalog contains 15,411 galaxies with
measured red shifts across 84\% of the sky. The starting point of this catalog
was the Infrared Astronomical Satellite (IRAS) Point Source Catalog (PSC). The
depth of the PSC is approximately 0.6 Jy. To translate redshift $z$ into
distance, we use Hubble's law $d = cz/H_0$ where $c$ is the speed of light and
$H_0 = 71~\mbox{km s}~^{-1}~\mbox{Mpc}~^{-1}$. 

We create a volume-limited sample (PSCz VL hereafter) by selecting PSCz entries
with $1~\mbox{Mpc}< d < 60~\mbox{Mpc}$ and $S_{60}d^2 > (0.6~\mbox{Jy})
(60~\mbox{Mpc})^2$, where $S_{60}$ is the flux at $60~\mu\mbox{m}$. Members of
the Local Group are excluded. The PSCz VL has 1329 galaxies. This corresponds to
a number density in the absence of clustering of $2 \times
10^{-3}~\mbox{Mpc}^{-3}$. The number of galaxies in the PSCz VL with $d <
10~\mbox{Mpc}$ is $2 \times$ larger than the number expected in the absence of
clustering. 

If the distribution of CR* sources is similar to the PSCz VL, then this
local over-density may be an important feature.
A local over-density of sources creates a greater probability
for a few nearby (and thereby bright) sources, even though the total number of
sources may be relatively high. Thus, a local over-density means greater number of source pairs than would
otherwise be expected.

The nearest galaxy in the PSCz VL is IC342, a Sc galaxy with
starburst activity. Our estimate of its distance using recessional velocity is
3.2 Mpc. From the luminosity of Cepheids, IC342 is 3.3 Mpc distant with a
luminosity $M_B \approx -20.7$ \citep{Karachentsev}.

\subsection{Further details of the Monte Carlo algorithm} \label{sec:OM}

We assume each source accelerates protons with an injection spectrum $dn/dE
\propto E^{-2.6}$ and a maximum energy $E_{max} = 3 \times 10^{20}~\mbox{eV}$.
The choice of injection spectrum does not strongly affect our results. 
The choice of particle type only influences the horizon at which CR* sources can be observed.
Changing the particle type to iron does not strongly affect this horizon
because protons and iron nuclei have similar energy loss lengths at this energy.

We assume that the observed flux of each source is constant over the observation period (e.g., years).
It should be noted that the observed lifetime of a CR* burst will be significantly lengthened due to particles taking different
paths from the source to earth.
Considering magnetic deflections in the galactic disk on the order of a few degrees, we should expect the shortest
CR* burst to be observed over a period of approximately 1 year.

We take
into account energy losses due to inelastic interactions with background
radiation fields (the GZK effect). We do this by using the continuous energy
loss approximation. We do not consider energy losses due to the expansion of the
universe. For the propagation distances we consider, these redshift losses are
negligible. Our test volume is a sphere centered at earth with radius $D =
250~\mbox{Mpc}$. We have checked that increasing $D$ does not change our
results. 

The angular distribution of CR* from each source (i.e., the surface brightness
of the source) is modeled as an elliptical Gaussian function with $\sigma_x =
\sigma_y = \bar{s}/2$. This is not an approximation, but it is
simply how we have chosen to define effective angular size.
Because we are
only interested in the value of $s$ averaged over the brightest sources (i.e.,
$\bar{s}$), it is not necessary to model how $s$ changes over different regions
of the sky or with source distance.
%
%
We assume that
the centroid of the Gaussian function is at the source location. As demonstrated
in Section~\ref{sec:morphology}, the centroid is expected to be offset from the
source location because of the regular component of the galactic magnetic field.
However, because $m$ depends only on the relative directions of the CR*,
neglecting this offset does not introduce a bias. 

It is appropriate to consider that a real cosmic-ray observatory has limited sky
coverage. Ultra-high energy cosmic-ray observatories that use a ground array
typically cover a large declination range with no small-scale structure
in their sky coverage. For example, their sky coverage is well approximated
by the function given by \citet{Sommers}.
For cases like this, other details of the sky coverage
(e.g., the exact declination limits) have little impact on our results.
Therefore, instead of simulating the sky coverage for a specific observatory, we
simulate the most general case, an observatory with equal coverage to all parts
of the sky.

\subsection{Generating a CR* Data Set}

To generate a single CR* data set, we randomly disperse sources with a number
density $\rho$ throughout the test volume according to one of our source
distribution models (evenly distributed or PSCz VL). Each CR* in a data set of
size $n_{obs}$ is randomly associated with a source. The probability that a CR*
is
associated with a given source is proportional to the expected flux of the
source, where the expected flux is a function of distance only. The CR*
directions are randomly disbursed from their source directions as described in
Section~\ref{sec:OM}. We test three different source sizes: $\bar{s} = 2.5^\circ$,
$\bar{s} = 5^\circ$, and $\bar{s} = 10^\circ$.

We generate CR* data sets with either $n_{obs}=92$ or $n_{obs}=184$. These
values of $n_{obs}$
corresponds to the number of CR* expected to be observed by the Pierre Auger
Observatory \citep{Abraham2004} at its fully deployed southern site
over a 4 and 8 year time span \citep{Abraham2007}, respectively. The Pierre
Auger Observatory is expected to reach $n_{obs}=92$ in the year 2011, and
$n_{obs}=184$ in
the year 2015.

After a simulated event set is generated, the value of $m$ is calculated with
Eq.~(\ref{eq:ClusteringMetric}). 
We calculate the expected range of $m$ for a given set of input parameters by
running 1000 Monte Carlo simulations.
This process is repeated for different values of $n_{obs}$, $\rho$ and
$\bar{s}$.


\section{Results and discussion}

In Fig.~\ref{fig:Results1}, we show our results with four graphs. The upper two
graphs are for $n_{obs} = 92$. The lower two graphs are for $n_{obs} = 184$. The
left two
graphs are for the PSCz VL source model. The right two graphs are for the evenly
distributed source model. Each graph shows the expected range of $m$ for
different $\bar{s}$ and $\rho$ scenarios. The error bars represent 10-90\%
quantiles.

\begin{figure}[t]
\vspace*{2mm}
\begin{center}
\includegraphics[width=8.3cm]{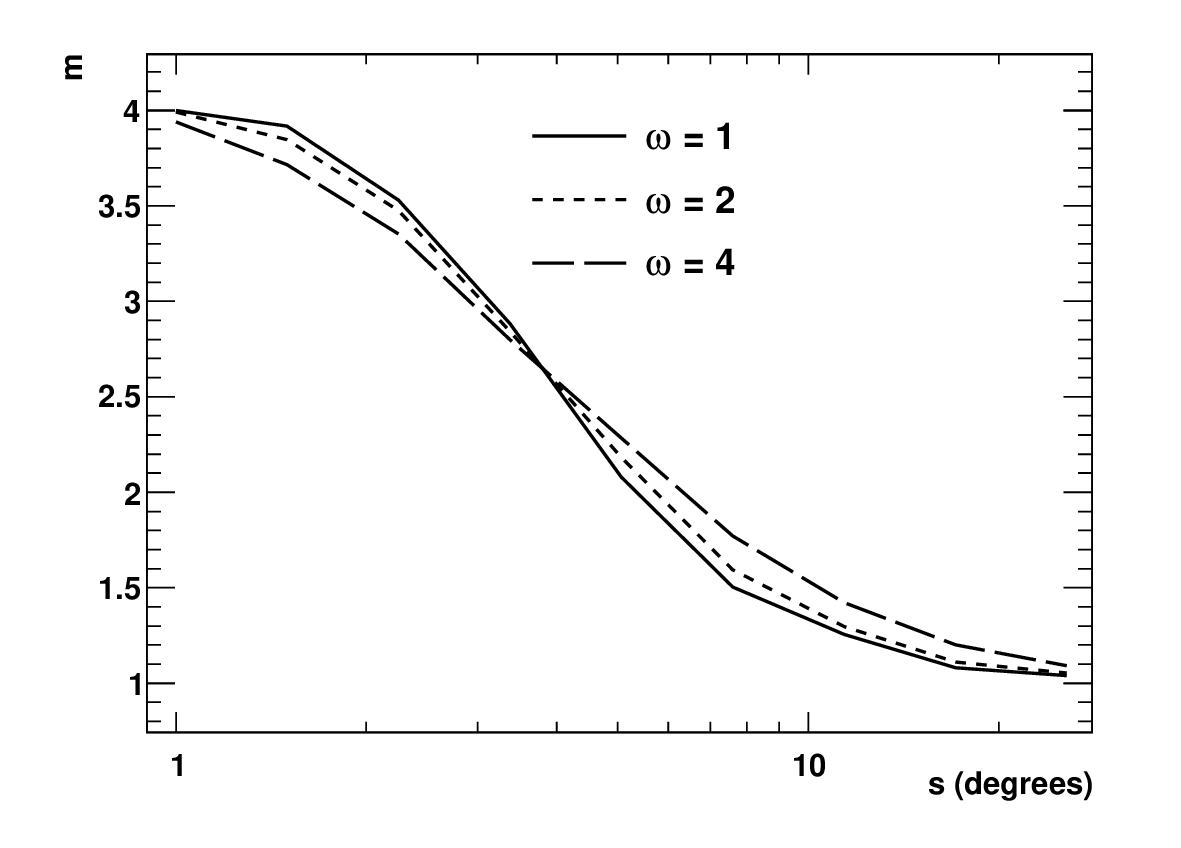}
\end{center}
\caption{Clustering metric $m$ as a function of
$s$ and $\omega$ from a single elliptical Gaussian source when $n_{obs}$ is large.}
\label{fig:AspectRatio}
\end{figure}

\begin{figure*}[t]
\vspace*{2mm}
\begin{center}
\includegraphics[width=18cm]{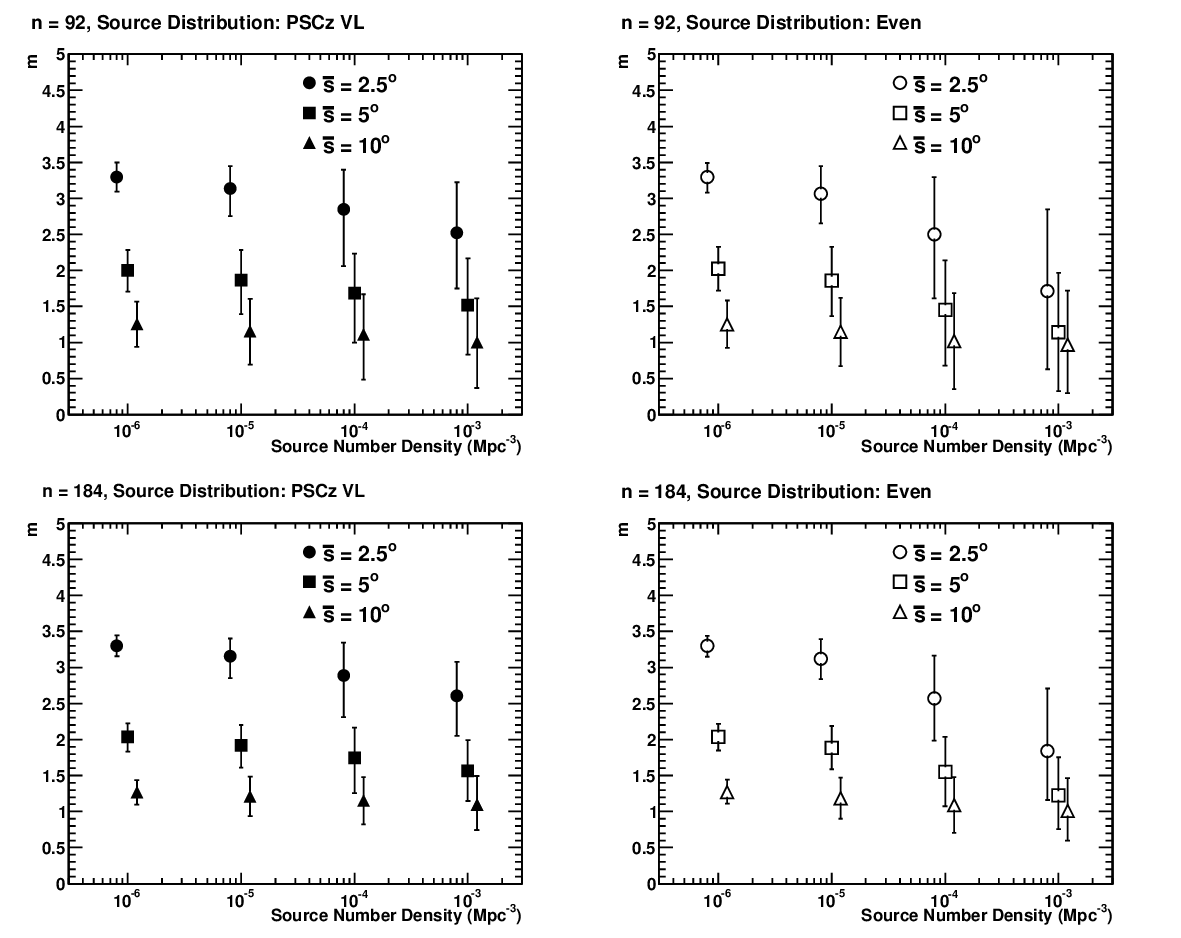}
\end{center}
\caption{Expected values of $m$ as a function of $n_{obs}$, $\bar{s}$, and
$\rho$. The markers are slightly offset from each other on the x-axis for clarity. The
error bars represent the 10-90\% quantile range.}
\label{fig:Results1}
\end{figure*}

The results shown in Fig.~\ref{fig:Results1} have the following general trends.
As $\rho$ increases or $n_{obs}$ decreases, the value of $m$ moves toward 1.
This occurs because the number of source pairs approaches zero.
As $\rho$ decreases or $n_{obs}$ increases, the value of $m$ moves toward the
value given in Fig.~\ref{fig:AspectRatio}. (The value of $m$ asymptotically
approaches a number somewhat less than the value given in
Fig.~\ref{fig:AspectRatio} when $\rho$ is large so that the angular spacing
between
sources is less than $10^\circ$.) This occurs because the
number of source pairs becomes a large number.
The number of source pairs is proportional to $n_{obs}^{2}$ and approximately 
proportional to $\bar{Q} \propto \rho^{-1/3}$~\citep{Younk}.

For source number densities $\rho \leq 10^{-4}~\mbox{Mpc}^{-3}$, the expected
range of $m$ is similar for the two source distribution
models. This shows our results are not strongly dependent on the details of the source
distribution if $\rho \leq 10^{-4}~\mbox{Mpc}^{-3}$.

For source number densities $\rho = 10^{-3}~\mbox{Mpc}^{-3}$ and for $\bar{s} = 2.5^\circ$,
the evenly
distributed model predicts a markedly smaller value for $m$ compared to
the PSCzVL model. 
The model detail that creates this difference is the local over-density
of sources.

The main conclusion from Fig.~\ref{fig:Results1} is the following. 
The metric $m$ is an effective discriminator of different $\bar{s}$ scenarios.
The discrimination power is best when $\rho$ is small. If we assume $\rho =
10^{-6}~\mbox{Mpc}^{-3}$, $m$ can easily differentiate between our three
$\bar{s}$ scenarios even with only $n_{obs}=92$ CR*. For example, $m = 2$
would favor $\bar{s} = 5^\circ$ and would disfavor both $\bar{s} = 2.5^\circ$
and $\bar{s} = 10^\circ$. This conclusion is independent of the source
distribution model. 

Simple checks like the above example will be an
important test for models that purport a small-scale angular correlation between
CR* and a set of astronomical objects. For example, consider a CR* source model
where the sources are a certain class of active galaxies with a number density $\rho =
10^{-6}~\mbox{Mpc}^{-3}$, the CR* are protons subject to the GZK effect, and the
CR* arrival directions are disbursed $2.5^\circ$ from the source. Then by
definition, $\bar{s} \leq 2.5^\circ$ ($\ll 2.5^\circ$ if the CR* are deflected
coherently). If we make the conservative assumption of no local over-density, we
must expect $m > 3.0$ when $n_{obs}=92$. If this value of $m$ is not
observed, the CR* source model cannot be considered self-consistent. That is,
for this scenario, it is rare not to find several CR* pairs in the
data set that are separated by less than $2.5^\circ$.

If we relax our constraint on $\rho$, the discrimination power of $m$ decreases
but is still meaningful. If we assume $10^{-6}~\mbox{Mpc}^{-3} \leq \rho \leq
10^{-3}~\mbox{Mpc}^{-3}$, with $n_{obs}=92$, the clustering metric $m$ can
differentiate
between $\bar{s} = 10^\circ$ and $\bar{s} = 5^\circ$. For example, $m = 2$ would
favor either $\bar{s} = 5^\circ$ or $\bar{s} = 2.5^\circ$  and would disfavor
$\bar{s} = 10^\circ$. Again, this conclusion is independent of the source
distribution model. 

\begin{figure}[t]
\vspace*{2mm}
\begin{center}
\includegraphics[width=8.3cm]{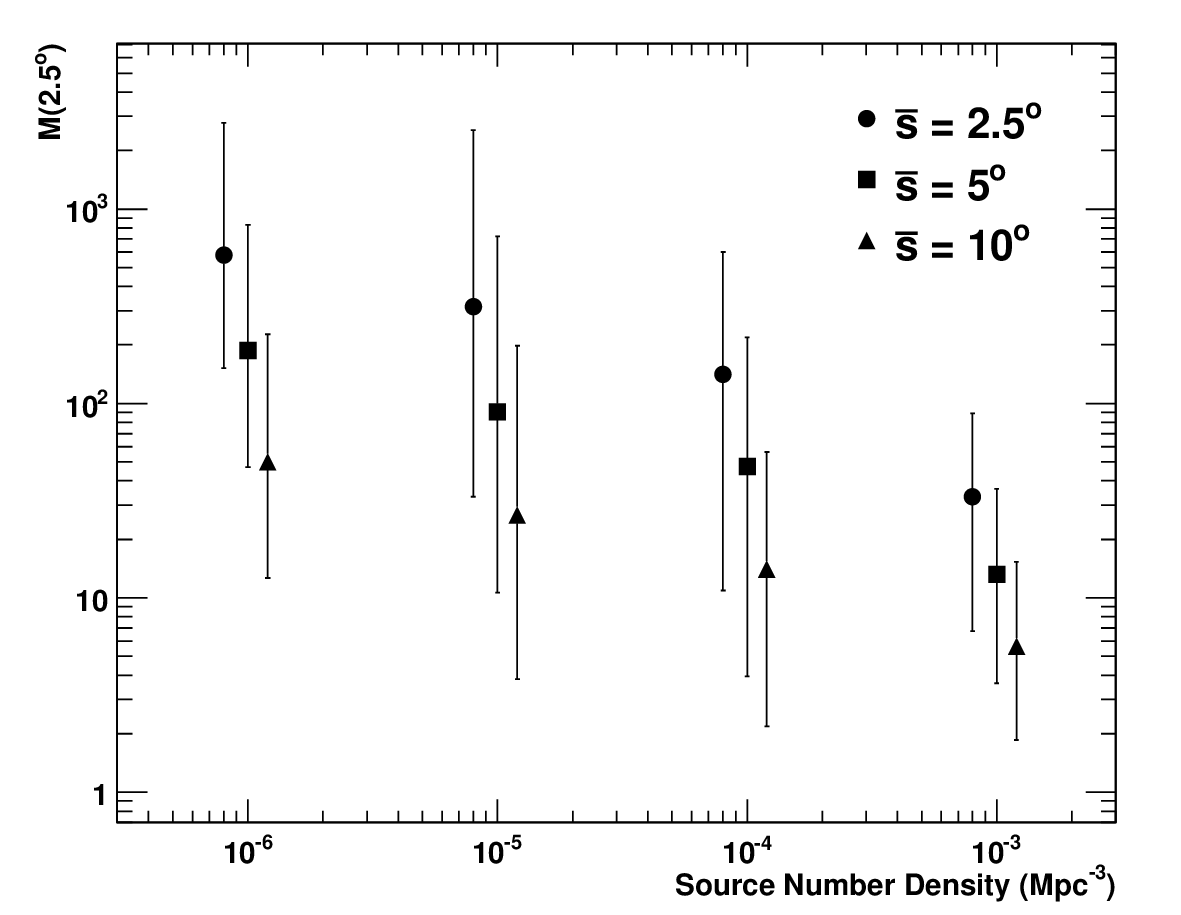}
\end{center}
\caption{Expected values of $M(2.5^\circ)$ as a function of $\bar{s}$ and
$\rho$, with $n_{obs} = 184$ and the PSC VL source distribution. The markers are slightly offset from each other on the x-axis for clarity. The
error bars represent the 10-90\% quantile range.}
\label{fig:Compare}
\end{figure}

Our ability to constrain $\bar{s}$ is
a direct result of how we defined $m$.
If we would have defined our clustering metric as the 
absolute amount of clustering at a single particular small angle
(e.g., $M(2.5^\circ)$),
our constraints would not be as powerful.
This was discussed in Section 3.
To demonstrate this, we show in Fig.~\ref{fig:Compare} the
expected values of $M(2.5^\circ)$ as a function of $\bar{s}$ and
$\rho$, with $n_{obs} = 184$ and the PSC VL source distribution.
By comparison with Fig. 3, it is clear that $M(2.5^\circ)$ is less telling of the value of $\bar{s}$ than $m$.
Although not shown, the same is true for other angles (e.g., $M(10^\circ)$) and whether or not
in the calculation of $M$
the pairs are weighted by the inverse of their angular separation.
Thus, in regards to constraining $\bar{s}$, there is a significant advantage to defining the clustering metric
in a way similar to Eq 1.

It is interesting to consider the situations where $\bar{s} < 10^\circ$ is
clearly favored. When a data set of 92 CR* has a clustering metric $m > 1.7$, or
a data set of 184 CR* has a clustering metric $m > 1.5$, we can conclude $\bar{s}
< 10^\circ$ with 90\% confidence. These constraints are not a strong function of
$\rho$ or of the source distribution model. 

If $\bar{s} < 10^\circ$ is indeed found to be favored, the simplest
interpretation is that the CR* are protons and the magnetic deflection is
similar to or possibly slightly greater than that predicted by the
model of the galactic magnetic field described by \citet{Harari1999}. In this instance, it is
not likely that the CR* are Helium nuclei because the energy loss length of
these particles is only a few Mpc \citep{Harari2006}. Also, it is not likely
that the CR* are more highly charged nuclei because their magnetic rigidity and
our knowledge of the magnetic field in the thin disk implies a magnetic
deflection that is difficult to reconcile with $\bar{s} < 10^\circ$. Thus, an
observation of a significantly large $m$ will constrain the model for magnetic
deflection and favor the idea 
that a significant fraction of CR* are protons.

If $\bar{s} \geq 10^\circ$ cannot be ruled out in the near future, there are
three possible interpretations. The first interpretation is that $\bar{s}$ is
actually small but $\rho$ is very large and there are no nearby sources.
This situation would delay the appearance of source pairs.
The second
interpretation is that $\bar{s} \geq 10^\circ$ because magnetic fields in the
thick disk or in extragalactic space deflect protons significantly more than the
magnetic fields in the thin disk. For example, this has been suggested by \citet{Ryu}.
The third interpretation is that $\bar{s} \geq
10^\circ$ because the CR* are heavy nuclei like iron ($Z \approx 26$). Indeed,
an iron-like composition at the highest energies is indicated by observations
reported by \citet{Abraham2010b}. This indication is not certain because it is
not currently possible to decouple ultra-high energy composition measurements
from the phenomenology of high energy particle interactions. In this context, a
constraint on $\bar{s}$ can also be used to constrain the phenomenology of high
energy particle interactions.

We limited this study to one definition of CR*, cosmic-rays with energy $E_{th}
> 57 \times 10^{18}~\mbox{eV}$. We believe this is the niche energy where the
brightest sources will stand out strongly from a background of dimmer sources, 
where source pairs  will exist in data sets of the near future, and where it is
plausible that the source size is small (i.e., $\bar{s} < 10^\circ$).
We consider $57 \times 10^{18}~\mbox{eV}$ a round number because of its use as 
a threshold energy by \citet{Abraham2007}.
It will be useful to consider other values of
$E_{th}$, although this is beyond the scope of this work. To take into account
the finite energy resolution and energy biases of a real cosmic-ray observatory,
testing other values of $E_{th}$ is required.

\conclusions[Summary]

We introduce a method to constrain the characteristic angular size of the
CR* sources under the general assumptions that the CR sources are extragalactic 
and that the GZK effect is operational.

We presented predictions of the clustering metric $m$,
as a function of $\bar{s}$, $n_{obs}$, and the distribution of sources.
We showed how, in the near future, an observed value of the clustering metric can constrain
the value of $\bar{s}$. A discrete source of CR* does not need to be identified,
or more generally, a strong localization of CR* directions does not need to be observed.
For example, the absence of small-scale clustering can be used to constrain $\bar{s}$.
We showed that constraints on $\bar{s}$ can be made rather independent of the assumed
spatial distribution and luminosity function of the cosmic-ray sources.
We must emphasize that any such constraints are dependent on the validity of our starting
assumptions and simplifications delineated in Section 4.

Constraints on $\bar{s}$ will be telling of the magnetic deflection of CR* and
the sources of CR*.
Differentiating between the two scenarios $\bar{s} < 10^\circ$ and $\bar{s} \geq 10^\circ$
will be particularly useful.





\end{document}